# Research on Composite Bit Technology for Hard Formations and Its Application in Igneous Rock

CHEN Lian[1, *], ZHAO Jiayuan[1], WEI Xiaohu[1], SONG Zhaohui[2], YANG Liyuan[3], ZHU Jintao[2]

1 School of Mechatronic Engineering, Southwest Petroleum University, Chengdu, Sichuan, 610500, China;
2 CNPC Xibu Drilling Engineering Co., Ltd, Urumchi, Xinjiang, 830000, China;
3 International Engineering Company, CNPC Chuanqing Drilling Engineering Co., Ltd, Chengdu, Sichuan, 610051, China

**Abstract**: The igneous rocks in deep formation have the characteristics of hardness, poor drillability and high abrasiveness, which is a difficulty in speeding up drilling. The drilling efficiency of existing conventional bits is low in igneous rocks. Based on the characteristics of igneous rocks, rock mechanical parameters and drillability experiments of granite, sandstone and other rocks were carried out. The rock drilling experiments of composite bit, tri-cone bit and PDC bit were carried out. Experiments have shown that in granite with very high strength, the drilling efficiency of conventional cone bit is very low, and it is extremely difficult for PDC bit to penetrate. The impact crushing effect of the cone of the composite bit can make the rock at the bottom of the well produce pits and cracks, which can assist the PDC cutters to penetrate into the formation, and solve the problem of the PDC cutters difficulty in penetrating in hard formations. In softer formations, the rock-breaking advantage of composite bit is not obvious, and the rock-breaking efficiency is lower than that of PDC bit. However, in hard formations, the advantage of composite bit is obvious, with higher drilling efficiency than PDC bit and cone bits. The personalized composite bit developed for deep igneous rocks formations has fast drilling speed, strong sustained drilling ability, long footage, and significant drilling speed-up effect. It significantly reduces the number of runs in deep drilling operations and achieves good application results. The composite bit is suitable for drilling in deep igneous hard-to-drill formations, and it has obvious advantages in deep igneous formations. It is a good choice for drilling speed-up in this kind of hard-to-drill formation.

## Introduction

The rocks in deep formation have the characteristics of hardness, poor drillability and high abrasiveness, which is a difficulty in speeding up drilling [1-3]. Igneous rock (also known as magmatic rock) formations are typical hard formations that are difficult to drill [4]. Igneous rock commonly includes granite, basalt, andesite, rhyolite, etc. Those types of hard formations are the difficulty in drilling and rock breaking operations. Existing conventional bits generally have low drilling efficiency and limited speed-up effect when drilling these high hardness igneous formations [5-6]. The Cone bit breaks the rock by impact crushing, mainly overcoming the compressive strength and hardness of the rock to make the rock broken. In hard formations, cone bits are inefficient in rock-breaking and slow in rate of penetration (ROP). Because of the limitation of the service life of the cone bit bearing sealing system, the footage of the cone bit is short and the drilling efficiency is low. The PDC bit breaks the rock by scraping and cutting. The precondition for efficient rock breaking is that the PDC cutters can effectively penetrate into the formation. In hard formations, PDC cutters are difficult to penetrate into the rock, resulting in low drilling efficiency. Especially in hard formations of igneous rocks, PDC cutters are easy to abrasion and collapse, which makes it difficult to adapt hard formations, resulting in short service life and less footage of PDC bit [7].

The composite bit is composed of cone and PDC fixed cutting structure [8]. In hard formations, interbedded layers, gravel layers and other complex hard-to-drill formations, the composite bit has obvious advantage. The composite bit is suitable for special drilling conditions that require a high control and stability of the tool surface [9-11]. This investigation analyzes the basic principle of composite bit. In accordance with the rock mechanics properties of hard formations, the authors carried out research on composite bit technology. In addition, the authors carried out the development and application of products related to composite bit, which provide reference for the development of composite bit technology and engineering application.

## 1. Structure and working principle of composite bit

The composite bit is constituted of the PDC fixed cutting structure and the cone non-fixed cutting structure to form the composite cutting structure [12]. The composite bit mainly adopts PDC scraping to break rock, and the cone rolling stamping to auxiliary break rock [13]. As shown in Fig. 1, the working principle of the composite bit is that driven by the rotation of the bit body, the cone makes the rock at the bottom of the well produce pits and cracks, and the PDC cutter scrapes the crushed rock in the bottom of the well.

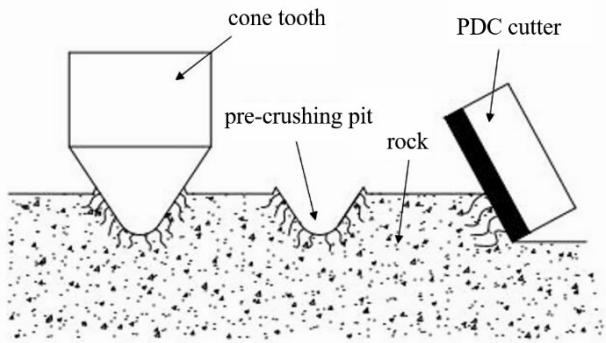

**Fig. 1. Rock breaking principle of compound drill bit**

The impact crushing effect of the cone make the rock at the bottom of the well produce pits and cracks, which can assist the PDC cutters to penetrate into the formation, and solve the problem of the PDC cutters difficulty to penetrate in hard formations. At the same time, the impact crushing effect of the cone makes the rock produce pits and cracks, which can weaken strength of the rock, reduce the scraping load and power consumption of the PDC cutters, and facilitate the scraping of PDC cutters. In addition, the uneven bottom of the well makes the continuous scraping of the PDC cutters become intermittent scraping, which reduce the thermal wear of PDC cutters. It is beneficial for the cooling of PDC cutters and prolongs the service life of PDC cutters. Therefore, the composite cutting structure not only improves the rock-breaking efficiency, but also prolongs the service life and footage of the bit. In the hard formations, the advantage of composite bit to speeding up drilling is obvious, and these effects are more prominent.

## 2. Rock Mechanics and Drillability Test Analysis

Rock drillability is one of the important indexes for measuring the difficulty of rock breakage, and is an important reference for drilling engineering design [14-15], such as bit optimization design, drilling speed-up process parameters optimization, etc. In accordance with the hard formation characteristics of igneous rocks, this investigation chose extremely hard granite for rock mechanical parameters and drillability test (Fig. 2). Meanwhile, the investigation chose the harder limestone and the softer sandstone to carry out comparative experiments (Table 1).

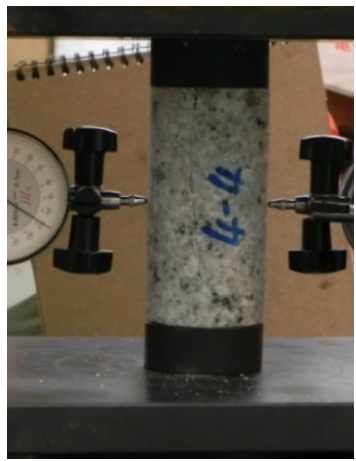

(a) uniaxial compressive strength test

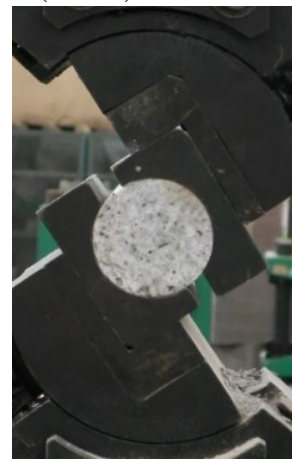

(b) shear strength test

**Fig. 2. Experimental testing of mechanical parameters of granite**

**Table 1 Rock mechanical parameters and drillability level**

| rock | uniaxial compressive strength (MPa) | Tensile strength (MPa) | Shear strength (MPa) | Internal friction angle(°) | drillability level cone bit | PDC bit |
|---|---|---|---|---|---|---|
| granite | 126.52 | 6.678 | 13.70 | 45.29 | 9.89 | ＞10 |
| limestone | 105.95 | 6.758 | 17.72 | 43.62 | 6.66 | 7.01 |
| sandstone | 67.55 | 4.346 | 13.56 | 38.03 | 5.76 | 5.48 |

In accordance with the experimental tests, the compressive strength and the angle of internal friction of granite were significantly higher than those of limestone and sandstone. Especially for the drillability level of the granite, cone bit drillability level is close to 10, and PDC bit has drillability level higher than the maximum level of 10 specified in the drillability standard [16]. This means that the granite has very high strength, so the drilling efficiency of conventional cone bit on the granite is very low, and the PDC bit is extremely difficult to penetrate.

## 3. Laboratory experimental study of composite bit

### 3.1 Indoor drilling experiment

As shown in Fig. 3, the authors designed and manufactured an experimental prototype machine for composite bit, and carried out indoor rock drilling experiments with composite bit, tri-cone bit and PDC bit on granite. Fig. 4 shows the granite bottomhole drilled by the composite bit.

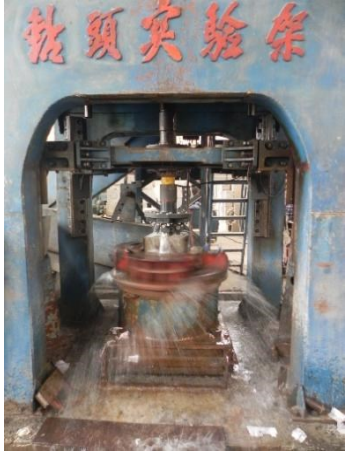

Fig. 3. Experimental prototype of composite bit

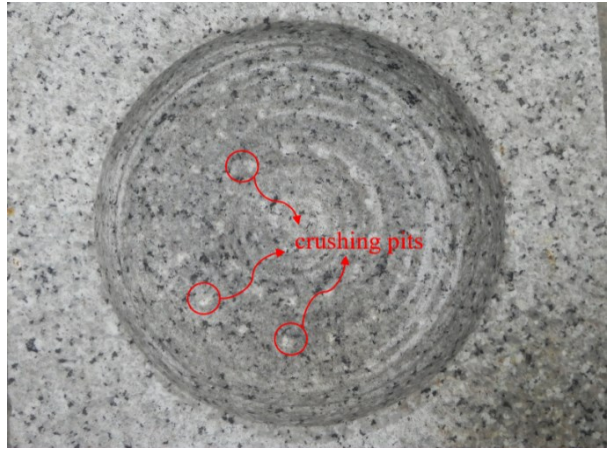

Fig. 4. Granite bottom drilled by composite bit

### 3.2 Experimental result analysis

As shown in Fig. 4, The bottom-hole pattern of the composite bit is obviously different from that of the tri-cone bit and the PDC bit. The cone teeth of the composite bit form punctiform distributed pits in the bottom-hole by crushing. While PDC cutters form circular tracks in the bottom-hole by scraping. Under the combined action of PDC cutters and cone teeth, the overlapped area of cones and fixed blades is shaped complex and rugged bottom-hole with circular cutting tracks and crushing pits. The rock at the bottom-hole in the common overlapped area is broken by the combined action of the two rock breaking methods mentioned above. The scraping of the PDC cutters and the impact crushing of the cone teeth combined action on the rock at the bottom-hole. The crushing pits formed by the cone teeth make the bottom of the well become rugged, which not only improve PDC cutters to penetrate into the rock, but also weaken rock strength, reduce cutting power consumption of PDC cutters, and beneficial for PDC cutters cooling.

As show in Fig. 5, it describes the correlation curve of the weight on bit (WOB)-ROP relationship for PDC bit, composite bit, and tri-cone bit drilling three different types of rocks. The ROP and WOB of three kinds of bits are positive correlation, when the WOB applied during drilling of three rocks in a certain range. When drilling softer sandstone (Fig. 5a), the PDC bit rely on the excellent cutting performance of PDC cutters that the ROP is much higher than cone bits and composite bits. Under the same WOB, because the load required for the cone teeth to penetrate rock is obviously higher than PDC cutters, the rock-breaking efficiency of the composite bit in soft formations is lower than that of the PDC bit. Therefore, in softer formations, the rock-breaking advantage of composite bit is not obvious. When drilling high hardness granite (Fig. 5c), the ROP of composite bit is higher than that of PDC bit and cone bit.

In the practical drilling, the main reason for the low efficiency of PDC bits in breaking hard rock formations is that the high hardness of the rock makes it difficult for cutters to effectively penetrate the rock [17]. Hard rock is easy to cause impact damage to PDC cutters [18]. Due to the high contact stress of the cutter edge, the thermal wear of the cutter is easy to occur [19]. These problems result in slow drilling speed, short footage, frequent tripping, and low drilling efficiency of PDC bits in hard formations. The composite bit combines the advantages of PDC bit and cone bit cutting structures. The pits stamping by the cone teeth can assist the PDC cutters in penetrating into the rock, which solves the problem of difficulty in PDC cutters penetrating into hard formations like granite. Meanwhile, the impact crushing effect of the cone makes the rock produce pits and cracks, which can weaken strength of the rock, improve the efficiency of PDC cutters in scraping and breaking rocks. So, compared to conventional PDC bits and cone bits, the composite bits have higher ROP and rock breaking efficiency when drilling hard formations. This agrees with the results of the previous analysis of rock drillability test, further indicates that the composite bit has significant advantage in breaking rocks when drilling hard formations.

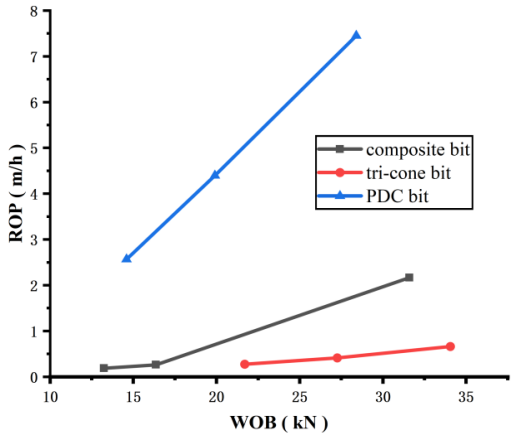

(a) limestone

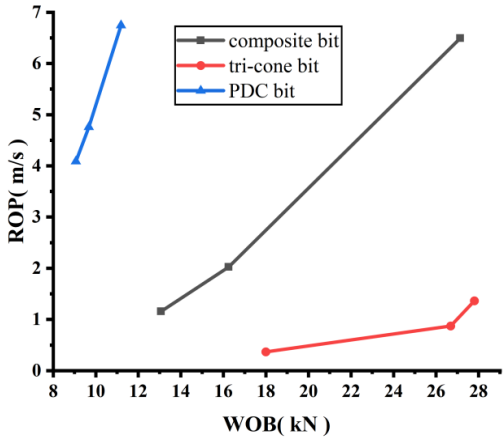

(b) sandstone

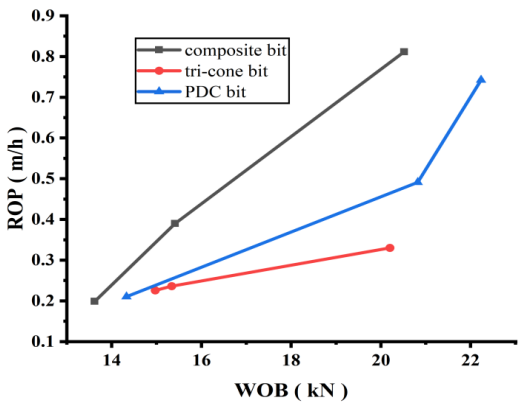

(c) granite

**Fig. 5 Correlation curve of the WOB-ROP relationship for three types of rocks using different bits**

As shown in Fig. 6, the authors plotted contrast curves of the WOB-torque relationship for PDC bits, composite bits, and cone bits when drilling three different rocks respectively. As seen in Fig. 6, Whether drilling softer sandstone or extremely hard granite, PDC bits have greater torque, followed by composite bits, and cone bits have the lowest torque. This is because PDC cutters mainly break rock through shearing and squeezing. During the process of shearing rock breaking, the interaction area between PDC cutters and rock is large, the cutting load is high, and the bit generates larger torque. Even though the PDC cutters have difficulty penetrating relatively hard granite, the shearing rock breaking method will still cause relatively large torque to be generated on the bit. As shown in Fig. 5, when drilling granite with high hardness, the torque of the composite bit is lower than that of the PDC bit, while its ROP is higher than that of the PDC bit. This further illustrates that composite bits fully utilize the cutting advantages of PDC bits and have significant rock breaking effects in hard formations. The composite bit also combines the low torque characteristics of the cone bit, which can significantly reduce the bit torque while ensuring high ROP, reduce the lateral vibration and stick-slip trend of the bit, improve the rock breaking ability and guided drilling ability of the bit, and have obvious advantages in directional drilling.

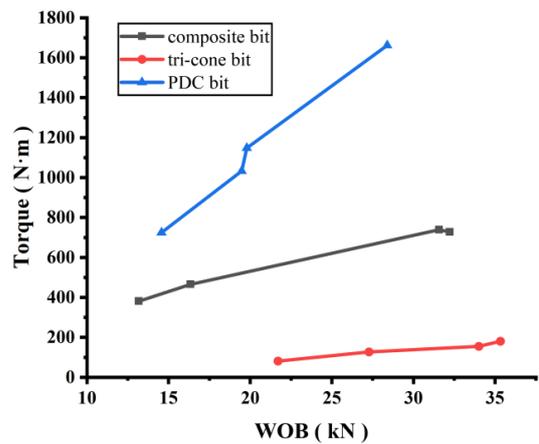

(a) limestone

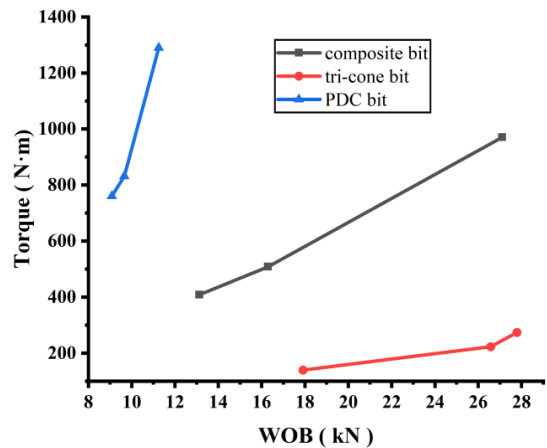

(b) sandstone

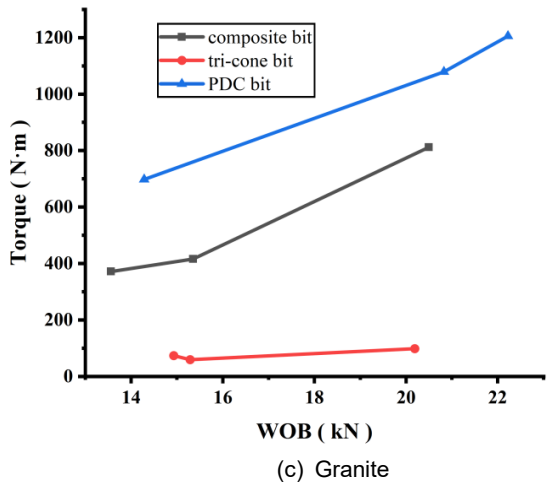

(c) Granite

**Fig. 6 Correlation curve of the WOB-Torque relationship for three types of rocks using different bits**

## 4. Design of composite bit for hard formations

The PDC cutting structures and cone cutting structures of the composite bit jointly undertake the rock breaking task. Because these two cutting structures play their respective roles while being influenced by each other, it is necessary to arrange their relative positions reasonably in order to fully utilize the advantages of the two cutting structures and minimize the negative effects between them. On the basis of researching the rock breaking mechanism of composite bits, the authors conducted research on the cutting structure design method of composite bits for hard formations in igneous rocks, laying the foundation for the development of subsequent bit products.

### 4.1 Cutting structure matching scheme

In the design of composite bits, the matching design of cutting structures is the most critical and important part, which directly determines the drilling efficiency and service life of the composite bit. According to the size of the composite bit, the matching scheme for the cutting structure of composite bits generally is constructed with two cutting blades and two cones (2+2 type, commonly used for small-sized bits), three cutting blades and three cones (3+3 type, commonly used for large-sized bits). At the same time, it is necessary to ensure that cones and cutting blades are evenly spaced in the circumferential direction [20-22].

Based on the characteristics of hard formations in igneous rocks and the goal of drilling speed-up, the designed composite bit is of the type with four cutting blades two cones (4+2 type). These four PDC fixed cutting blades are distributed in an "X" shape, where the combination of two blades and a single cone is symmetrically arranged in the circumferential direction (Fig. 7). The 4+2 type composite bit increases the number of PDC fixed cutters, which not only maximizes the cutting advantages of PDC cutting structure and improves ROP, but also increases the space for the fixed gauge protection structure of the bit, increases the supporting area of the bit at the bottom of the well, and further enhances the stability of the bit operation. Because of the low rock-breaking efficiency of the cone, and the excessive number of cones sharing more energy, the ROP of the bit is low. Two cones can fully utilize the load buffering and pre-crushing effects of the cones, reduce the load impact of PDC cutters, and enhance the efficient and sustainable drilling capability of the bit.

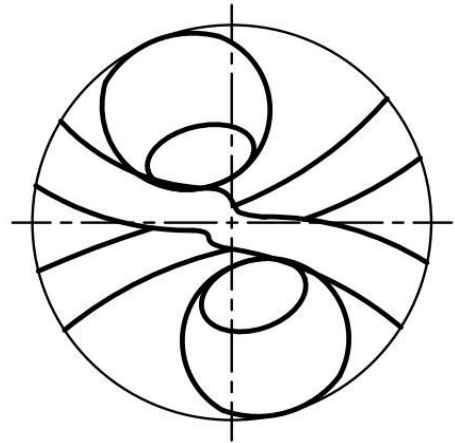

**Fig. 7 Designed 4+2 type composite bit**

### 4.1 Contour coordination design

The design of contour coordination needs to focus on the stability of bit operation, the coordination of cone cutting structure, and the adaptability to drilling technology. To ensure the rational contour design of the two cutting structures of composite bits, it is necessary to first determine the primary and secondary relationship between the two cutting contours [23]. In the hard formation, the cone cutting structure of the composite bit mainly assists the PDC cutting structure in rock breaking, improves the ability of the PDC cutting structure to penetrate into the formation, and prolongs the service life of PDC cutting structure. Therefore, first design the contour of the PDC cutting structure.

The cutting contour line of the PDC bit is also known as the cutter crown curve. The shape of the cutter crown not only has an impact on the stability, drilling speed, bottom hole cleaning effect, and service life of the bit, but also has different adaptability to the formation depending on the shapes of the crown. At the same time, it can also determine the size of the bit cutter distributing surface [24]. As shown in Fig. 8, The cutting contour line (crown curve) of PDC bit is mainly composed of the inner cone, nose, shoulder, and gauge protection part. The International Association of Drilling Contractors (IADC) categorizes crown curve profile shapes into four basic types: flat cone type, shallow cone type, medium cone type, and long cone type [25]. Based on the characteristics of hard formations in igneous rocks, the crown shape of the composite bit adopts medium cone type. The medium cone type crown has a certain depth of inner cone and a large cutter distributing area, which ensures good stability of the bit. It can also be equipped with more cutting cutters to make the wear of the bit more uniform, improve the rock breaking efficiency and service life of the bit. When drilling in these hard formations of igneous rocks, the medium cone type crown shape concentrates the WOB on the top of the bit, and the cutting cutters on the top of the crown can produce more effective shear action on the rock.

The cones adopt a radial continuous coverage design in the radial direction. The continuous coverage design enhances the role of the cones in assisting rock breaking under complex motion conditions during composite drilling of the bit, and improves the ROP. Meanwhile, the design satisfies the matching relationship between the cone teeth distributing and the PDC fixed blades cutters distributing. In addition, since the 4+2 composite bit has four fixed blades, the blades occupy a large space, especially in the core area of the bit. To provide sufficient space for placing the central nozzle in the core area, the blades shape in the core area is made into a reverse circular arc shape, as shown in Fig. 9.

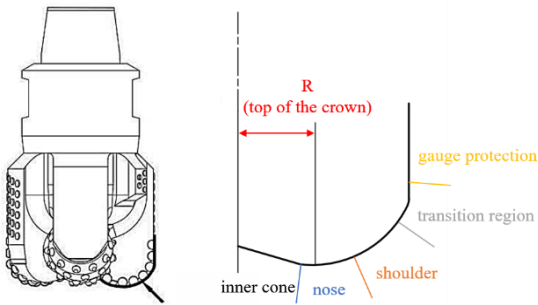

**Fig. 8 Components of PDC bit crown curve**

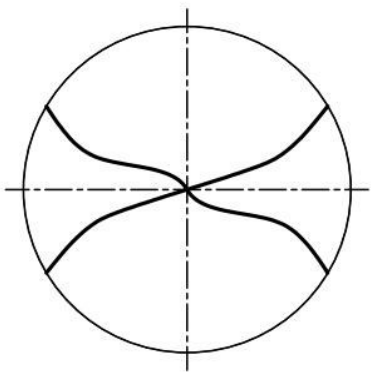

**Fig. 9 Anti circular arc-shaped blade**

## 5. Field applications of composite bits in the hard formation

### 5.1 Engineering problem and bit development

The above research indicates that composite bits rely on the stamping effect of the cones to assist PDC cutters in penetrating the formation, solving the problem of the difficulty for PDC cutters to penetrate hard formations. The composite cutting structure not only can improve rock breaking efficiency, but also prolong the service life and footage of the bit.

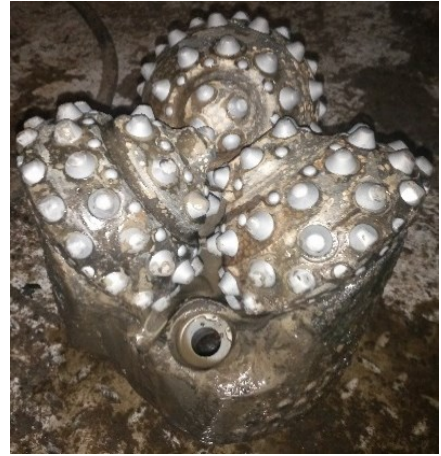

**Fig. 10 Cone bit used in the Archean**

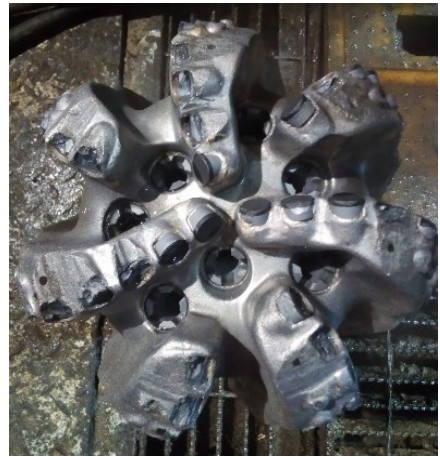

**Fig. 11 Damaged PDC bit from the Archean**

The rocks in the deep formation of the Archean (3500-4500m) in the Liaohe Oilfield of China are mainly migmatitic granite. The rock has extremely high hardness, high compressive strength, and strong abrasiveness. The formation has poor drillability. As shown in Fig. 10, the cone bits are prone to teeth dull grinding, slow ROP. Due to the limited lifespan of bearings, the footage of cone bits is short. As shown in Fig. 11, the PDC bits are prone to damage when drilling these hard formations, which leads to a decrease in the footage per bit and slow ROP. Therefore, conventional bits in this formation have low drilling efficiency, resulting in high costs. The composite bits have obvious advantages in drilling speed-up in hard formations, with strong continuous drilling ability and long service life. The application of composite bits to drill the migmatitic granite formation is an effective way to solve the above problems. As shown in Fig. 12, based on the geological characteristics of this section the migmatitic granite formation in the Archean, a composite bit has been developed. The bit adopts a composite cutting structure design of four cutting blades and two cones (4+2 type), with a balanced design of blades and cones, resulting in high stability of the bit.

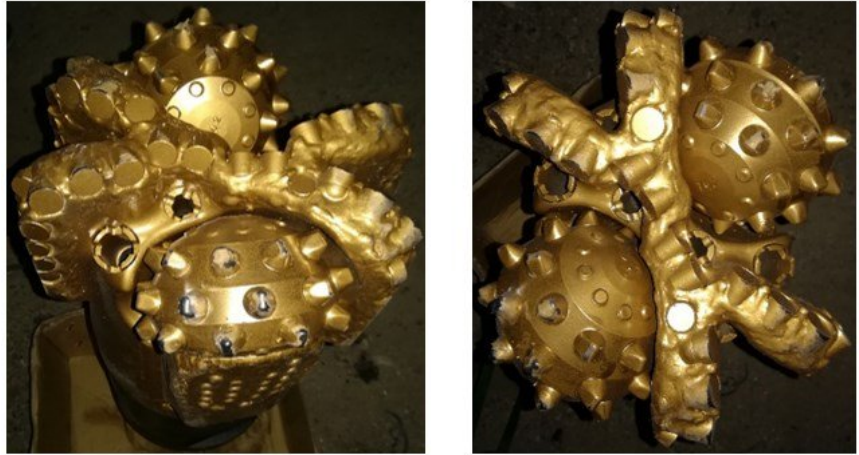

**Fig. 12 Developed 4+2 type composite bit**

## 5.2 Field applications of composite bits

The 8-½" composite bit (H2416) developed for deep granite formations was used in three drilling runs in the Archean granite interval of the Xinggu area in the Liaohe Oilfield. The Archean granite section of Xinggu 7-A well is 680m (3978-4658m) long, and a total of eight drilling runs were used in this section, with the third runs using a composite bit and the rest using PDC bits (Table 2). The composite bit drilled for 43 hours, with a footage of 155m (4080-4235m) and a ROP of 3.60m/h.

**Table 2 Application of composite bits in granite formations**

| well number | position | lithology | bit type | well-depth (m) | footage (m) | drilling time (h) | ROP (m/h) |
|---|---|---|---|---|---|---|---|
| Xinggu 7-A | Archean | granite | PDC bit (P616HG) | 3978-4021.28 | 43.28 | 16.89 | 2.56 |
| | | | PDC bit (M713BG) | 4021.28-4080 | 59 | 16.7 | 3.53 |
| | | | composite bit (H2416) | 4080-4235 | 155 | 43 | 3.60 |
| | | | PDC bit (M713BG) | 4235-4328 | 93 | 24.73 | 3.76 |
| | | | PDC bit (M716H) | 4328-4418 | 90 | 23.89 | 3.77 |
| | | | PDC bit (M716H) | 4418-4503.54 | 85.54 | 28.36 | 3.02 |
| | | | PDC bit (M716H) | 4503.54-4585.8 | 82.26 | 25 | 3.29 |
| | | | PDC bit (M716H) | 4585.8-4658 | 72.2 | 18 | 4.01 |
| Xinggu 7-B | Archean | granite | composite bit (H2416) | 4165-4390 | 225 | 54.9 | 4.1 |
| | | | composite bit (H2416) | 4390-4595 | 205 | 45.1 | 4.5 |

| | | | | |
|---|---|---|---|---|
| PDC bit (P616G) | 4595-4672 | 77 | 28.8 | 2.7 |
| PDC bit (P716H) | 4672-4798 | 126 | 35.7 | 3.5 |

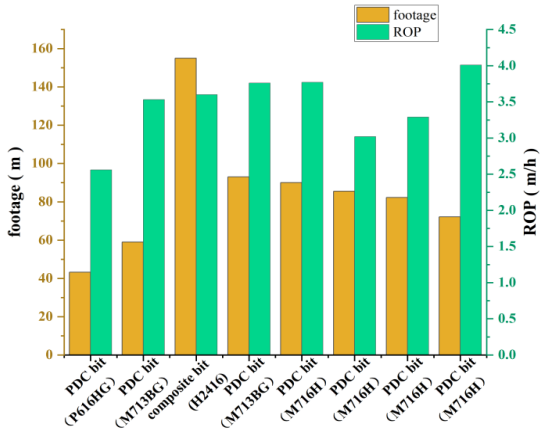

Fig. 13 Comparison curve of drilling indicators for Xinggu 7-A well bits

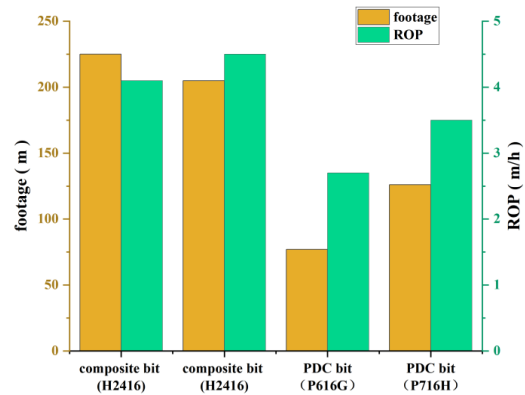

Fig. 14 Comparison curve of drilling indicators for Xinggu 7-B well bits

The Archean granite section of Xinggu 7-B well is 634m long (4165-4798m), and a total of four drilling runs were used in this section, with the first two runs using composite bit and the last two runs using PDC bits (Table 2). The footage of two composite bits both exceeded 200m (205m and 225m), with ROPs of 4.1m/h and 4.5m/h, respectively. The footage of two PDC bits is 77m and 126m respectively. As shown in Fig. 14, in the granite formation of the well, the performance of the composite bits is also obviously better than that of the PDC bits. The footage of composite bits is 106.6-111.8% higher than that of PDC bits, the ROP is over 17.1% higher than that of PDC bits, and the lifespan of composite bits is 26.3-90.6% longer than that of PDC bits. Fig. 15 shows the last composite bit from the well. It can be seen from the figure that the composite bit is in good condition and not severely worn out after drilling. Most of the PDC cutters wore less than 1mm, and there are three PDC cutters on the outer shoulder with wore of about 2mm, and one PDC cutter is broken. The wore of the cone teeth is about 1.5mm.

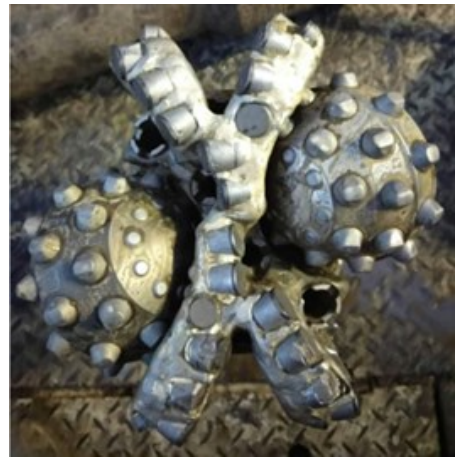

Fig. 15 Composite bit after drilling

Field applications have shown that composite bits have strong continuous drilling capabilities. The composite bits with PDC cutting structures and cones can adapt well to drilling in deep granite formations that are difficult to drill. Based on the laboratory experiment results mentioned above, the field applications once again prove that the impact crushing effect of the cone causes crushing pits at the bottom of the well and damages the rock (such as cracks), weakening the rock strength. This function can assist PDC cutters in penetrating the formation, thereby improving the rock breaking efficiency and ROP of the bit. The effect of cone making pits and weakening rocks obviously reduces the cutting load and power consumption of PDC cutters, reduces impact, protects PDC cutters, reduces damage, prolongs the service life of bits, increases footage, and reduces the number of drilling runs.

Therefore, the composite bit has obvious advantages in deep igneous hard drilling formations. It is a good choice for drilling speed-up in this kind of hard-to-drill formation.

# 6. Conclusions

(1) The rock mechanics properties and drillability tests indicate that in granite with very high strength, the drilling efficiency of conventional cone bits is very low, and it is extremely difficult for PDC bits to penetrate. The bottomhole pattern of the composite bit indicates that the impact crushing effect of the cone causes pits and cracks in the rock at the bottom of the well, which can assist the PDC cutter's in penetrating the formation and solve the problem of difficult penetration of PDC cutters in hard formations.

(2) In softer formations, the rock-breaking advantage of composite bits is not obvious, and the rock-breaking efficiency is lower than that of PDC bits. However, in hard formations, the advantage of composite bits is obvious, with higher drilling efficiency than that of PDC bits and cone bits. Composite bits are suitable for drilling in hard formations.

(3) Field applications show that composite bits can adapt well to drilling in difficult to drill granite formations in deep. The role of cones in making pits and weakening rocks can assist PDC cutters in penetrating the formation, improve the efficiency of rock breaking and ROP of the bit, significantly reduce the cutting load and power consumption of PDC cutters, reduce impact, protect PDC cutters, reduce damage, extend the service life of the bit, increase footage, and reduce the number of drilling runs. The composite bits are a good choice for speeding up drilling in deep igneous hard drilling formations.

# Acknowledgements

This work was supported by National Natural Science Foundation of China (51304168), and the China Post-doctoral Science Foundation (Grant No. 2021M693909). Supported by High Performance Computing Center, Southwest Petroleum University.